\documentclass[aip,jap,reprint,groupedaddress]{revtex4-1}%
\usepackage{natmove}
\usepackage{graphicx}
\usepackage{epstopdf}
\usepackage{dcolumn}
\usepackage{bm}
\usepackage{color}
\usepackage[hidelinks]{hyperref}
\usepackage{subfigure}
\usepackage{amsmath}%
\usepackage{amsfonts}%
\usepackage{amssymb}
\hypersetup{
    colorlinks=true,       
    linkcolor=blue,
    citecolor=blue,       
    filecolor=blue,      
    urlcolor=blue            
}

\begin{document}
\title{\textcolor{blue}%
{Density matrix Monte Carlo modeling of quantum cascade lasers}}
\author{Christian Jirauschek}
\email{jirauschek@tum.de}
\homepage{http://www.cph.ei.tum.de}
\affiliation
{Department of Electrical and Computer Engineering, Technical University of Munich (TUM), D-80333 Munich,
Germany}
\date{16 November 2017, published as J. Appl. Phys. 122, 133105 (2017)}
\begin{abstract}
By including elements of the density matrix formalism, the semiclassical ensemble Monte Carlo method for carrier transport is extended to incorporate incoherent tunneling, known to play an important role in quantum cascade lasers (QCLs). In particular, this effect dominates electron transport across thick injection barriers, which are frequently used in terahertz QCL designs. A self-consistent model for quantum mechanical dephasing is implemented, eliminating the need for empirical simulation parameters. Our modeling approach is validated against available experimental data for different types of terahertz QCL designs.
\end{abstract}
\maketitle

\section{\label{sec:level1}Introduction}

Advanced carrier transport modeling techniques for semiconductor devices
evaluate the relevant processes, such as different scattering mechanisms,
directly based on the corresponding Hamiltonians. Consequently, these
approaches do not require specific experimental or empirical input, but only
rely on well known material parameters. Especially for the modeling of
advanced semiconductor nanodevices, theoretical methods beyond the standard
drift-diffusion model are required. Various theoretical approaches with
different levels of complexity are available, such as the ensemble Monte Carlo
(EMC) method \cite{MC} or the nonequilibrium Green's function (NEGF) approach
\cite{Schwinger,kadanoff_book, Keldysh_1965,jirauschek2014modeling}.

Here, we focus on the modeling of quantum cascade lasers (QCLs)
\cite{1994Sci...264..553F}. These devices are highly interesting from a
scientific point of view as well as with respect to various applications,
e.g., in metrology and sensing, and are already commercially available
\cite{2010CPL...487....1C}. The active region consists of a quantum well
structure, where the laser levels are formed between the quantized electron
states in the conduction band. Thus, the lasing wavelength does not primarily
depend on the material system used, but can be selected over a wide spectral
range by adequate quantum design. In particular, the mid-infrared and
terahertz (THz) regions become accessible, thus complementing the spectral
range covered by conventional semiconductor lasers.

Modeling provides detailed insight into the complex interplay of the physical
effects involved and allows for a systematic optimization of QCL designs,
e.g., with respect to efficiency, operating temperature and spectral range.
The ideal simulation approach should offer excellent accuracy and versatility,
combined with decent computational efficiency to allow for design optimization
over an extended parameter range. Advanced carrier transport modeling
techniques not only consider the quantized energy states due to the electron
confinement in growth direction, but also the in-plane wavevectors which are
related to the kinetic electron energies. In this way, both the inter- and
intrasubband carrier dynamics can be considered. EMC has been widely employed
for the analysis and design of QCLs, offering a good compromise between
reliability and numerical effort
\cite{borowik2017monte,2000ApPhL..76.2265I,2001ApPhL..79.3920K,2003ApPhL..83..207C,2004ApPhL..84..645C,2002ApPhL..80..920C,2006ApPhL..88f1119L,2007JAP...101f3101G,2007JAP...102k3107G,matyas2010temperature,matyas2011photon,2009JAP...105l3102J,2005JAP....97d3702B}%
. However, as a semiclassical method EMC neglects quantum coherence effects,
most notably tunnel coupling between the two states spanning the injection
barriers and the level broadening of the states \cite{2009PhRvB..79p5322W}.
Such effects become relevant especially for thick barriers, where the states
involved in the electron transport have a small energy separation, and the
electron transport is governed by resonant tunneling. This applies especially
to the injection barriers in various types of THz QCL designs
\cite{2005JAP....98j4505C,2009PhRvB..79p5322W}. Consequently, also quantum
transport approaches are frequently used for the simulation of QCLs, including
NEGF
\cite{2002PhRvB..66h5326W,banit_wacker,2009PhRvB..79s5323K,2009ApPhL..95w1111S,Kubis_assess,Schmielau_ktyp,Kolek_openQCL,2012ApPhL.101u1113W,Wacker_JSTQE,grange2015contrasting}
as well as the density matrix (DM)
\cite{2001PhRvL..87n6603I,2009PhRvB..79p5322W,iotti2016electronic,lindskog2014comparative}
and related Wigner function \cite{jonasson2015dissipative} formalism. The
computational load of these approaches is much higher than for their
semiclassical counterparts, impeding their applicability to complex QCL
structures with many subbands, and to QCL design and optimization in general.

Various strategies exist to obtain a simulation approach which includes the
most important quantum effects and still provides acceptable computational
efficiency. Particularly for the DM approach, the in-plane wavevector
dependence is frequently neglected, reducing the order of the density matrix
to the number of considered subbands
\cite{2009PhRvB..80x5316K,2010PhRvB..81t5311D,2010NJPh...12c3045T,dinh2012extended,fathololoumi2012terahertz}%
. Another approach is to reduce the numerical burden of full quantum transport
simulation methods by introducing simplifications. In particular, for NEGF
various approximations have been developed which greatly simplify the
evaluation of the scattering self-energies. This includes the constant $k$
approximation which assumes that the scattering self-energies are independent
of the in-plane wavevector \cite{Schmielau_ktyp,Kubis_assess,Wacker_JSTQE},
and the multi-scattering B\"{u}ttiker probe model where the lesser
self-energies are replaced by a quasi-equilibrium expression
\cite{greck2015efficient}. An opposite strategy is to start with a
semiclassical method such as EMC and extend it to include quantum effects
relevant for the carrier transport. An example is the consideration of
collisional broadening in EMC \cite{2009JAP...105h3722A,2013ApPhL.102a1101M}.
Furthermore, the DM formalism has been combined with EMC to describe the
incoherent tunneling transport across thick injection barriers
\cite{2005JAP....98j4505C,2012ApPhL.100a1108B,freeman2016self,freeman2012nonequilibrium}%
. This hybrid approach overcomes the main weakness of semiclassical transport
simulations, which treat the carrier transport through a barrier as
instantaneous since electrons scattered into states extending over the barrier
see no resistance \cite{2005JAP....98j4505C}. For biases where narrow
anticrossings occur, this can result in excessive current spikes, indicating
the breakdown of the semiclassical description \cite{2010PhyE...42.2628M}.

In this paper, we extend EMC to include incoherent tunneling transport,
building upon the pioneering work of Callebaut and Hu
\cite{2005JAP....98j4505C}. As in previous related approaches
\cite{2005JAP....98j4505C,2012ApPhL.100a1108B,freeman2016self,freeman2012nonequilibrium}%
, we use localized basis states to describe the incoherent tunneling
mechanism. However, here this effect is not considered by solving the von
Neumann and the Boltzmann transport equations in parallel, but rather directly
built into EMC as an additional ''scattering-like'' mechanism derived from the
DM\ formalism. This enables a straightforward implementation into existing EMC
codes, without significantly increasing the numerical load. In this way, all
the features of advanced EMC QCL simulation tools, such as the consideration
of electron-electron scattering beyond the Hartree approximation and coupling
between carrier transport and optical cavity field
\cite{1988PhRvB..37.2578G,2010ApPhL..96a1103J,jirauschek2014modeling}, are taken advantage of.
Furthermore, in the resulting DM-EMC approach the self-consistent character of
EMC is preserved by implementing a self-consistent model for the dephasing
rate \cite{unuma2003intersubband,freeman2016self,ando1985line}, thus avoiding
the need for empirical parameters.

\section{\label{sec:level2}Method}

\subsection{Localized states and tunneling\label{sec:tun}}

\begin{figure}[ptb]
\includegraphics{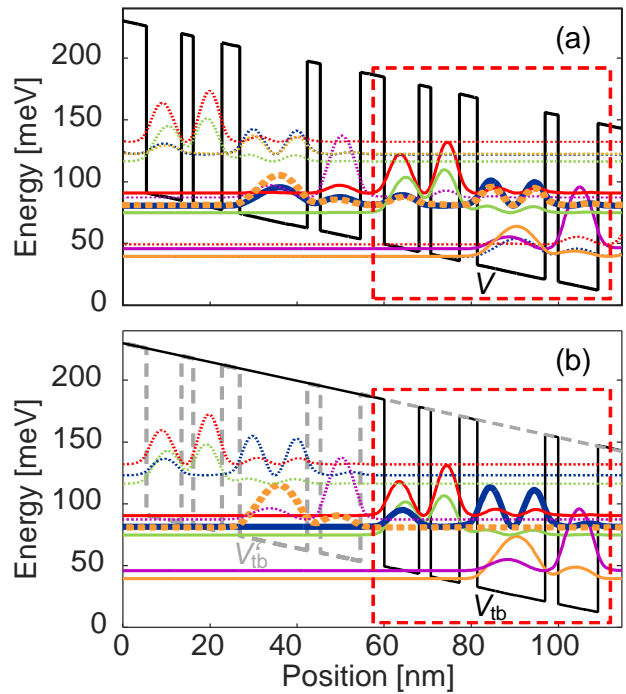}
\caption{Conduction band profile and probability densities for a four-well THz
QCL \cite{benz:apl:2007} at a bias of $7.6\,\mathrm{kV}/\mathrm{cm}$, computed
based on (a) the actual potential $V$ and (b) the tight-binding potential
$V_{\mathrm{tb}}$. The rectangles denote a single QCL period. The
wavefunctions involved in the electron transport across the thick barrier are
marked by bold lines.}%
\label{fig1}%
\end{figure}

Electron transport across thick barriers, such as the injection barriers in
various types of THz QCL designs, is governed by incoherent tunneling between
near-resonant states \cite{2005JAP....98j4505C,2009PhRvB..79p5322W}. For
these, the use of localized wavefunctions has proven advantageous, as can be
obtained from a tight binding approach
\cite{2005JAP....98j4505C,2012ApPhL.100a1108B,freeman2016self,freeman2012nonequilibrium}%
. Rather than determining the extended eigenstates for the actual potential
$V$ as in semiclassical simulations [see Fig.\thinspace\ref{fig1}(a)], the
active region structure is divided into modules separated by thick barriers,
where each module is described by a separate potential $V_{\mathrm{tb}}$,\ as
indicated in Fig.\thinspace\ref{fig1}(b). The wavefunctions and eigenenergies
are then calculated for each module separately using a Schr\"{o}dinger-Poisson
solver. In this tight-binding framework, the transport within a module can be
implemented into EMC in the usual way by evaluating scattering-induced
transport, since quantum coherence effects only appear for transitions across
the module boundaries. We note that this approach can straightforwardly be
extended to structures containing more than one thick barrier per period
\cite{2010PhRvB..81t5311D}.

In our model, an electron state $\left|  i,\mathbf{k}\right\rangle $ is
characterized by its subband energy $E_{i}$, tight-binding wavefunction
$\psi_{i}\left(  z\right)  $ where $z$ denotes the position coordinate along
the growth direction, and in-plane wavevector $\mathbf{k}$. In the following,
we assume decoupling between the confinement in $z$ direction and in-plane
motion which strictly holds for infinite quantum wells
\cite{1989PhRvB..40.7714E}, implying $\mathbf{k}$\ independent wavefunctions
$\psi_{i}\left(  z\right)  $. The decoupling approximation also works
reasonably well for finite, not too narrow quantum wells, and is frequently
used for QCL modeling to reduce complexity \cite{savic2006electron}.
Furthermore, decoupling is maintained for an elementary treatment of
nonparabolicity, where an energy dependent effective mass is used to determine
the $\psi_{i}\left(  z\right)  $ and $E_{i}$, and subsequently, for each
subband $i$ an effective mass $m_{i}^{\ast}$ is computed describing the
electron dispersion relation at the corresponding subband bottom
\cite{1989PhRvB..40.7714E,jirauschek2014modeling}.

Tunneling from a state $\left|  i,\mathbf{k}\right\rangle $ to states $\left|
j,\mathbf{k}\right\rangle $ across a barrier can be described by a density
matrix equation combined with rate equation terms
\cite{2001PhRvL..87n6603I,2010NJPh...12c3045T},
\begin{subequations}
\label{eq:dm2}%
\begin{align}
\mathrm{d}_{t}\rho_{ii,\mathbf{k}} &  =\sum_{\left(  j,\mathbf{k}^{\prime
}\right)  \neq\left(  i,\mathbf{k}\right)  }r_{j,\mathbf{k}^{\prime
}\rightarrow i,\mathbf{k}}\rho_{jj,\mathbf{k}^{\prime}}-r_{i,\mathbf{k}}%
\rho_{ii,\mathbf{k}}\nonumber\\
&  +\sum_{j}\mathrm{i}\Omega_{ij}\left(  \rho_{ij,\mathbf{k}}-\rho
_{ji,\mathbf{k}}\right)  ,\label{eq:dm2a}\\
\mathrm{d}_{t}\rho_{ij,\mathbf{k}} &  =\sum_{n}\mathrm{i}\left(  \Omega
_{nj}\rho_{in,\mathbf{k}}-\Omega_{in}\rho_{nj,\mathbf{k}}\right)
-\mathrm{i}\omega_{ij}\rho_{ij,\mathbf{k}}-\gamma_{ij,\mathbf{k}}%
\rho_{ij,\mathbf{k}}.\label{eq:dm2b}%
\end{align}
The diagonal density matrix elements $\rho_{ii,\mathbf{k}}$ correspond to the
electron occupation of state $\left|  i,\mathbf{k}\right\rangle $, while the
off-diagonal elements $\rho_{ij,\mathbf{k}}$ are related to the coherences
between states $\left|  i,\mathbf{k}\right\rangle $\ and $\left|
j,\mathbf{k}\right\rangle $. The scattering rates $r_{j,\mathbf{k}^{\prime
}\rightarrow i,\mathbf{k}}$ from all possible states $\left|  j,\mathbf{k}%
^{\prime}\right\rangle $\ to $\left|  i,\mathbf{k}\right\rangle $ correspond
in our approach to the conventional EMC scattering rates, and $r_{i,\mathbf{k}%
}$ is the total scattering rate from state $\left|  i,\mathbf{k}\right\rangle
$ into all other states,%
\end{subequations}
\begin{equation}
r_{i,\mathbf{k}}=\sum_{\left(  j,\mathbf{k}^{\prime}\right)  \neq\left(
i,\mathbf{k}\right)  }r_{i,\mathbf{k}\rightarrow j,\mathbf{k}^{\prime}%
}.\label{rik}%
\end{equation}
The modeling of the dephasing rate $\gamma_{ij,\mathbf{k}}$ is discussed in
Section \ref{sec:deph}. Furthermore, $\omega_{ij}=\left(  E_{i}-E_{j}\right)
/\hbar$ denotes the resonance frequency between states $\left|  i,\mathbf{k}%
\right\rangle $\ and $\left|  j,\mathbf{k}\right\rangle $, which is
$\mathbf{k}$ independent for parabolic subbands with identical effective mass.
With the extended and tight-binding potentials $V$ and $V_{\mathrm{tb}}$
illustrated in Figs.\thinspace\ref{fig1}(a) and \ref{fig1}(b), respectively,
the tunnel coupling is approximately described by the matrix element
$\left\langle i,\mathbf{k}\right|  V-V_{\mathrm{tb}}\left|  j,\mathbf{k}%
^{\prime}\right\rangle $. This expression becomes $0$ for $\mathbf{k}%
\neq\mathbf{k}^{\prime}$ since $V$ and $V_{\mathrm{tb}}$ only depend on $z$,
and is given by $\hbar\Omega_{ij}$ for $\mathbf{k}=\mathbf{k}^{\prime}$, with
the Rabi frequency $\Omega_{ij}=\left\langle \psi_{i}\right|  V-V_{\mathrm{tb}%
}\left|  \psi_{j}\right\rangle /\hbar$. Thus, in our model given by
Eq.\thinspace(\ref{eq:dm2}), incoherent tunneling only occurs between states
with identical $\mathbf{k}$
\cite{2001PhRvL..87n6603I,2005JAP....98j4505C,wacker2001transport,wacker}.
Furthermore, since in our approach this effect is only considered across thick
barriers separating the individual modules, $\Omega_{ij}$ is nonzero only for
the state doublets spanning these barriers \cite{2010NJPh...12c3045T}.

In the following, interference effects between different tunneling transitions
involving the same subband $i$ to the left or $j$ to the right of the barrier
are neglected, i.e., only the terms containing $\Omega_{ij}$ are kept in the
sum of Eq.\thinspace(\ref{eq:dm2b}) \cite{razavipour2013indirectly}. The
stationary solution of Eq.\thinspace(\ref{eq:dm2}) is obtained by setting
$\mathrm{d}_{t}=0$, which yields%
\begin{align}
0  &  =\sum_{\left(  j,\mathbf{k}^{\prime}\right)  \neq\left(  i,\mathbf{k}%
\right)  }r_{j,\mathbf{k}^{\prime}\rightarrow i,\mathbf{k}}\rho_{jj,\mathbf{k}%
^{\prime}}-r_{i,\mathbf{k}}\rho_{ii,\mathbf{k}}\nonumber\\
&  +\sum_{j}\left(  \rho_{jj,\mathbf{k}}-\rho_{ii,\mathbf{k}}\right)
r_{i\rightarrow j,\mathbf{k}}^{\mathrm{t}}, \label{eq:dmrate}%
\end{align}
with the single-electron tunneling rate from $\left|  i,\mathbf{k}%
\right\rangle $ to $\left|  j,\mathbf{k}\right\rangle $ given by
\begin{equation}
r_{i\rightarrow j,\mathbf{k}}^{\mathrm{t}}=\frac{2\Omega_{ij}^{2}%
\gamma_{ij,\mathbf{k}}}{\omega_{ij}^{2}+\gamma_{ij,\mathbf{k}}^{2}}.
\label{rij}%
\end{equation}
Equation (\ref{eq:dmrate}) is compatible with the EMC framework, since the
tunneling rate given in Eq.\thinspace(\ref{rij}) can be straightforwardly
implemented as a pseudo-scattering mechanism. For calculating $\Omega_{ij}$,
we consider the asymmetry of the structure with respect to the tunneling
barrier, introduced by the module design and the applied bias, by using the
improved formula $\left(  \hbar\Omega_{ij}\right)  ^{2}=\left\langle \psi
_{i}\right|  V-V_{\mathrm{tb}}\left|  \psi_{j}\right\rangle \left\langle
\psi_{i}\right|  V-V_{\mathrm{tb}}^{\prime}\left|  \psi_{j}\right\rangle $.
Here, $V_{\mathrm{tb}}^{\prime}$ and $V_{\mathrm{tb}}$ are the tight-binding
potentials of the modules to the left and right of the tunneling barrier,
respectively [see Fig.\thinspace\ref{fig1}(b)]
\cite{grier2015coherent,yariv1985approximate}. Pauli blocking is accounted for
by considering the final state occupation probability $f_{j,\mathbf{k}}$ in
the form of an additional factor $\left(  1-f_{j,\mathbf{k}}\right)  $ in
Eq.\thinspace(\ref{rij}). This correction can be implemented into the DM-EMC
algorithm by using the rejection method \cite{1985Lugli_excl}. A time
dependent inclusion of light-matter interaction in Eq.\thinspace(\ref{eq:dm2})
leads to similar terms as for tunneling, with the Rabi frequency now depending
on the optical field strength. Within the rotating wave approximation, the
photon-induced transition rates can be derived in a similar way as
Eq.\thinspace(\ref{rij}), again featuring a Lorentzian energy dependence and
$\mathbf{k}$ conservation \cite{Boyd,2010ApPhL..96a1103J,jirauschek2010monte}.
These rates are then included in Eq.\thinspace(\ref{eq:dmrate}) in the same
way as $r_{i\rightarrow j,\mathbf{k}}^{\mathrm{t}}$, and a self-consistent
evaluation of optical transitions can be achieved by coupled carrier transport
and optical cavity field simulations
\cite{2010ApPhL..96a1103J,jirauschek2010monte}. As for Eq.\thinspace
(\ref{rij}), quantum interference effects between different optical
transitions sharing a common level are neglected in the derivation of the
photon-induced transition rates, as well as interferences between optical and
tunneling transitions. We note that the latter can lead to a splitting of the
gain spectrum into two lobes
\cite{2009PhRvB..80x5316K,2010PhRvB..81t5311D,tzenov2016time}, which is not
included in the model presented here.

As a consequence of $\mathbf{k}$ conservation for tunneling in Eq.\thinspace
(\ref{eq:dm2}), energy is not conserved for state doublets with $\omega
_{ij}\neq0$ \cite{2009PhRvB..80x5316K}. Energy conservation can be restored by
including higher order corrections which give rise to scattering-assisted
tunneling \cite{Willenberg_2003,wacker2001transport,terazzi2012transport}%
.\ Since only doublets with small energy difference $\hbar\left|  \omega
_{ij}\right|  $ contribute significantly to the tunneling transport across the
thick barriers, as can be seen from Eq.\thinspace(\ref{rij}), the assumption
of $\mathbf{k}$\ conservation involved in Eq.\thinspace(\ref{rij}) has been
proven to work well for QCLs
\cite{2001PhRvL..87n6603I,2005JAP....98j4505C,2012ApPhL.100a1108B}.

\subsection{Modeling of dephasing rates\label{sec:deph}}

The dephasing rates $\gamma_{ij,\mathbf{k}}$ in Eq.\thinspace(\ref{rij}) are
calculated based on Ando's model
\cite{ando1978broadening,ando1985line,unuma2003intersubband}, which is
compatible with EMC, and thus also with the DM-EMC approach envisaged here
\cite{freeman2016self}. Ando's model has already been successfully used for
QCL simulations based on simplified DM methods
\cite{terazzi2012transport,2010NJPh...12c3045T,2010PhRvB..81t5311D}, advanced
DM approaches accounting for the in-plane electron dynamics
\cite{lindskog2014comparative}, and NEGF
\cite{2009JAP...106f3115N,banit_wacker}. In Ando's approach, the dephasing
rate is given by \cite{ando1985line,unuma2003intersubband}%
\begin{equation}
\gamma_{ij,\mathbf{k}}=(\gamma_{i,\mathbf{k}}+\gamma_{j,\mathbf{k}}%
)/2+\gamma_{ij,\mathbf{k}}^{\prime}. \label{gijk}%
\end{equation}
The first term corresponds to the lifetime broadening already implemented in
our EMC approach \cite{2009JAP...105l3102J}, which is computed based on the
outscattering rate from state $\left|  i,\mathbf{k}\right\rangle $ into other
subbands,
\begin{equation}
\gamma_{i,\mathbf{k}}=\sum_{j\neq i}\sum_{\mathbf{k}^{\prime}}r_{i,\mathbf{k}%
\rightarrow j,\mathbf{k}^{\prime}}. \label{gik}%
\end{equation}
We note that Eq.\thinspace(\ref{gik}) deviates from $r_{i,\mathbf{k}}$ defined
in Eq.\thinspace(\ref{rik}) in that Eq.\thinspace(\ref{gik}) excludes
intrasubband scattering, which is in Eq.\thinspace(\ref{gijk}) accounted for
by the so-called pure dephasing contribution $\gamma_{ij,\mathbf{k}}^{\prime}%
$. For calculating the tunnel resonance linewidth, Eq.\thinspace(\ref{gik})
contains all scattering mechanisms considered in the rates $r_{j,\mathbf{k}%
^{\prime}\rightarrow i,\mathbf{k}}$ of Eq.\thinspace(\ref{eq:dm2}), i.e., the
tunneling rate itself is not included in Eq.\thinspace(\ref{gik}). Equation
(\ref{gijk}) can also be used to calculate the linewidth of optical
transitions, which are described by a density matrix approach similar to
Eq.\thinspace(\ref{eq:dm2}) \cite{Boyd}, with stationary rate solutions
analogous to Eq.\thinspace(\ref{rij})
\cite{2010ApPhL..96a1103J,jirauschek2010monte}. In this case, the lifetime
broadening computed with Eq.\thinspace(\ref{gik}) considers all scattering
contributions with the exception of the stimulated optical transition rates,
but including the tunneling rates given by Eq.\thinspace(\ref{rij}). As
mentioned in Section \ref{sec:tun}, our approach neglects coherences between
optical and tunneling transitions, which can for example lead to a splitting
of the gain spectrum into two lobes
\cite{2009PhRvB..80x5316K,2010PhRvB..81t5311D,tzenov2016time}.

In Eq.\thinspace(\ref{gijk}), intrasubband contributions to broadening are
accounted for by the pure dephasing rate $\gamma_{ij,\mathbf{k}}^{\prime}$,
given by \cite{ando1985line}
\begin{align}
\gamma_{ij,\mathbf{k}}^{\prime}  &  =\frac{\pi}{\hbar}\iint\mathrm{d}%
^{2}\mathbf{k}^{\prime}\Big\{N\left(  \mathbf{k}\right)  \left\langle \left|
\left\langle i\mathbf{k}^{\prime}\right|  V^{\prime}\left|  i\mathbf{k}%
\right\rangle -\left\langle j\mathbf{k}^{\prime}\right|  V^{\prime}\left|
j\mathbf{k}\right\rangle \right|  ^{2}\right\rangle \nonumber\\
&  \times\delta\left[  \varepsilon\left(  \mathbf{k}\right)  -\varepsilon
\left(  \mathbf{k}^{\prime}\right)  \pm\hbar\omega_{0}\right]  \Big\}.
\label{gamp}%
\end{align}
Here, $\hbar$ denotes the reduced Planck constant, $N(\mathbf{k})=S/\left(
2\pi\right)  ^{2}$ is the 2D density of states in $\mathbf{k}$ space with the
in-plane cross section area $S$, $V^{\prime}$ denotes the scattering
potential, $\varepsilon$ is the kinetic energy, and $\delta$ denotes the Dirac
delta function. Furthermore, $\hbar\omega_{0}$ corresponds to the longitudinal
optical (LO) phonon energy for phonon absorption (+) and emission (-), and is
set to $0$ for elastic scattering processes. Lastly, $\left\langle
\dots\right\rangle $ denotes statistical averaging over the distribution of
scatterers. In this approach, the total broadening is obtained as the sum of
each individual broadening mechanism \cite{unuma2003intersubband}. Equation
(\ref{gamp}) assumes parabolic subbands with identical effective masses, but
can be generalized to the nonparabolic case
\cite{unuma2003intersubband,terazzi2012transport}. Since we use Eq.\thinspace
(\ref{gamp}) for optical as well as tunneling transitions, we re-emphasize
that optical transitions within a module occur between energy eigenstates,
while tunneling is here described in the localized representation based on
eigenstates of the so-called pseudospin Hamiltonian \cite{grifoni1998driven},
which should however not affect the validity of Eq.\thinspace(\ref{gamp}).

For calculating $\gamma_{ij,\mathbf{k}}^{\prime}$, we only consider ionized
impurity and interface roughness scattering, since contributions due to
electron-electron and LO phonon scattering have been found to be negligible as
the corresponding matrix elements largely cancel in Eq.\thinspace(\ref{gamp})
\cite{freeman2016self}. While Ando's derivation of Eq.\thinspace(\ref{gamp})
explicitly excludes effects due to electron-electron interactions
\cite{ando1978broadening}, it could be theoretically shown that for
transitions between parabolic subbands with identical effective masses, the
lineshape is barely affected by electron-electron interactions beyond
intersubband electron-electron collisions
\cite{nikonov1997collective,waldmuller2004optical}. Also the contribution of
LO phonon scattering to pure dephasing in quantum well structures was
investigated in detail, and found negligible even at room temperature
\cite{unuma2001effects,unuma2003intersubband}. We have verified for selected
tunneling and optical transitions that the LO phonon contribution to pure
dephasing\cite{unuma2003intersubband,terazzi2012transport} is indeed
negligible for the QCL\ designs considered here, as further discussed in
Section \ref{sec:rates}. Assuming in-plane isotropy, we introduce $k=\left|
\mathbf{k}\right|  $, $k^{\prime}=\left|  \mathbf{k}^{\prime}\right|  $,
$\varepsilon=\hbar^{2}k^{2}/\left(  2m^{\ast}\right)  $ with effective mass
$m^{\ast}$, $\varepsilon^{\prime}=\hbar^{2}\left(  k^{\prime}\right)
^{2}/\left(  2m^{\ast}\right)  $, and\ the angle $\theta$ between $\mathbf{k}$
and $\mathbf{k}^{\prime}$. Setting $\omega_{0}=0$ for elastic processes,
Eq.\thinspace(\ref{gamp}) then becomes with $\left\langle i\mathbf{k}^{\prime
}\right|  V^{\prime}\left|  i\mathbf{k}\right\rangle =V^{\prime}\left(
i,k^{\prime},k,\theta\right)  $%
\begin{equation}
\gamma_{ij,\mathbf{k}}^{\prime}=\frac{Sm^{\ast}}{4\pi\hbar^{3}}\int_{0}^{2\pi
}\mathrm{d}\theta\left\langle \left|  V^{\prime}\left(  i,k,k,\theta\right)
-V^{\prime}\left(  j,k,k,\theta\right)  \right|  ^{2}\right\rangle .
\label{gamp2}%
\end{equation}

\subsubsection{Interface roughness scattering}

The matrix element for interface roughness scattering is with $\mathbf{q}%
=\mathbf{k}-\mathbf{k}^{\prime}$ given by \cite{jirauschek2014modeling}
\begin{equation}
\left\langle j\mathbf{k}^{\prime}\right|  V^{\prime}\left|  i\mathbf{k}%
\right\rangle =\pm\frac{V_{\mathrm{o}}}{S}\psi_{i}\left(  z_{0}\right)
\psi_{j}^{\ast}\left(  z_{0}\right)  \int\mathrm{d}^{2}\mathbf{r}\Delta\left(
\mathbf{r}\right)  \exp\left(  \mathrm{i}\mathbf{qr}\right)  , \label{Vif}%
\end{equation}
where $z_{0}$ denotes the average interface position and $\Delta\left(
\mathbf{r}\right)  $ is the local deviation of the interface as a function of
the in-plane coordinates $\mathbf{r=}\left[  x,y\right]  $. $V_{\mathrm{o}}$
denotes the band offset, and the ''$+$'' (''$-$'') sign corresponds to a
barrier (well) located at $z<z_{0}$. The interface roughness is typically
described by a Gaussian autocorrelation function \cite{1987ApPhL..51.1934S}%
\begin{align}
\left\langle \Delta\left(  \mathbf{r}\right)  \Delta\left(  \mathbf{r}%
^{\prime}\right)  \right\rangle  &  =\frac{1}{S}\int\Delta\left(
\mathbf{r}\right)  \Delta\left(  \mathbf{r+d}\right)  \mathrm{d}^{2}%
\mathbf{r}\nonumber\\
&  =\Delta^{2}\exp\left(  -\frac{\mathbf{d}^{2}}{\Lambda^{2}}\right)
\label{eq:corr}%
\end{align}
with $\mathbf{d}=\mathbf{r}^{\prime}-\mathbf{r}$, where $\Delta$ and $\Lambda$
denote the average root-mean-square roughness height and in-plane correlation
length, respectively. Using Eqs.\thinspace(\ref{Vif}) and (\ref{eq:corr}), we
obtain
\begin{align}
&  \left|  \left\langle i\mathbf{k}^{\prime}\right|  V^{\prime}\left|
i\mathbf{k}\right\rangle -\left\langle j\mathbf{k}^{\prime}\right|  V^{\prime
}\left|  j\mathbf{k}\right\rangle \right|  ^{2}\nonumber\\
&  =\frac{V_{\mathrm{o}}^{2}}{S^{2}}\left|  \int\mathrm{d}^{2}\mathbf{r}%
\Delta\left(  \mathbf{r}\right)  \exp\left(  \mathrm{i}\mathbf{qr}\right)
\right|  ^{2}\left[  \left|  \psi_{i}\left(  z_{0}\right)  \right|
^{2}-\left|  \psi_{j}\left(  z_{0}\right)  \right|  ^{2}\right]
^{2}\nonumber\\
&  =\frac{V_{\mathrm{o}}^{2}}{S}\pi\Delta^{2}\Lambda^{2}\exp\left(  -\frac
{1}{4}\Lambda^{2}\mathbf{q}^{2}\right)  \left[  \left|  \psi_{i}\left(
z_{0}\right)  \right|  ^{2}-\left|  \psi_{j}\left(  z_{0}\right)  \right|
^{2}\right]  ^{2}.
\end{align}
Additionally summing over all interfaces located at positions $z_{n}$,
Eq.\thinspace(\ref{gamp2}) becomes%
\begin{align}
\gamma_{ij,\mathbf{k}}^{\prime}  &  =\gamma_{ij,k}^{\prime}=\frac
{V_{\mathrm{o}}^{2}\Delta^{2}\Lambda^{2}m^{\ast}}{2\hbar^{3}}\sum_{n}\left[
\left|  \psi_{i}\left(  z_{n}\right)  \right|  ^{2}-\left|  \psi_{j}\left(
z_{n}\right)  \right|  ^{2}\right]  ^{2}\nonumber\\
&  \times\int_{0}^{\pi}\mathrm{d}\theta\exp\left(  -\frac{1}{4}\Lambda
^{2}q^{2}\right)  , \label{gif}%
\end{align}
with $q^{2}=2k^{2}\left(  1-\cos\theta\right)  $. The integral can be
evaluated analytically,%
\[
\int_{0}^{\pi}\mathrm{d}\theta\exp\left(  -\frac{1}{4}\Lambda^{2}q^{2}\right)
=\pi\exp\left(  -\frac{1}{2}\Lambda^{2}k^{2}\right)  \mathrm{I}_{0}\left(
\frac{1}{2}\Lambda^{2}k^{2}\right)  ,
\]
where $\mathrm{I}_{\nu}$ is the modified Bessel function of the first kind.

\subsubsection{Impurity scattering}

The matrix element for impurity scattering is given by%
\begin{equation}
\left\langle j\mathbf{k}^{\prime}\right|  V^{\prime}\left|  i\mathbf{k}%
\right\rangle =-\frac{e^{2}}{2\epsilon qS}\int_{-\infty}^{\infty}\psi
_{i}\left(  z\right)  \psi_{j}^{\ast}\left(  z\right)  \exp\left(  -q\left|
z-z^{\prime}\right|  \right)  \mathrm{d}z, \label{Vimp}%
\end{equation}
where $\epsilon$ is the permittivity, $z^{\prime}$ denotes the $z$ component
of the impurity position, and $e$ is the elementary charge. We then obtain%
\begin{equation}
\left\langle \left|  \left\langle i\mathbf{k}^{\prime}\right|  V^{\prime
}\left|  i\mathbf{k}\right\rangle -\left\langle j\mathbf{k}^{\prime}\right|
V^{\prime}\left|  j\mathbf{k}\right\rangle \right|  ^{2}\right\rangle
=\frac{e^{4}}{4\epsilon^{2}q^{2}S}f_{ij}\left(  q\right)  , \label{Vimpd}%
\end{equation}
where%
\[
f_{ij}\left(  q\right)  =F_{ii}\left(  q\right)  -2F_{ij}\left(  q\right)
+F_{jj}\left(  q\right)  ,
\]
with%
\begin{align}
F_{ij}\left(  q\right)   &  =\int\mathrm{d}z^{\prime}\,n_{\mathrm{D}}\left(
z^{\prime}\right)  \left[  \int_{-\infty}^{\infty}\left|  \psi_{i}\left(
z\right)  \right|  ^{2}\exp\left(  -q\left|  z-z^{\prime}\right|  \right)
\mathrm{d}z\right] \nonumber\\
&  \times\left[  \int_{-\infty}^{\infty}\left|  \psi_{j}\left(  z\right)
\right|  ^{2}\exp\left(  -q\left|  z-z^{\prime}\right|  \right)
\mathrm{d}z\right]  .
\end{align}
Here we have integrated over the doping concentration $n_{\mathrm{D}}\left(
z\right)  $ to include the effect of all ionized impurities. The pure
dephasing rate is with Eq.\thinspace(\ref{gamp2}) and $q\left(  \theta\right)
=2^{1/2}k\left(  1-\cos\theta\right)  ^{1/2}$ given by
\cite{unuma2003intersubband}%
\begin{equation}
\gamma_{ij,\mathbf{k}}^{\prime}=\gamma_{ij,k}^{\prime}=\frac{e^{4}m^{\ast}%
}{8\pi\hbar^{3}}\int_{0}^{\pi}\mathrm{d}\theta\frac{f_{ij}\left[
2^{1/2}k\left(  1-\cos\theta\right)  ^{1/2}\right]  }{2k^{2}\left(
1-\cos\theta\right)  \epsilon^{2}}. \label{gimp}%
\end{equation}
Screening can for example be considered in the random phase approximation
(RPA), resulting in modified matrix elements \cite{2006ApPhL..89u1115L}.
Simplifications of the RPA, such as the single subband screening model, result
in a description of screening in terms of a constant inverse screening length
$q_{\mathrm{s}}$, which is introduced by formally substituting $\epsilon
\rightarrow\epsilon\left(  1+q_{\mathrm{s}}/q\right)  $ in Eqs.\thinspace
(\ref{Vimp}), (\ref{Vimpd}) and (\ref{gimp}) \cite{2006ApPhL..89u1115L}.

\section{\label{sec:level3}Results and discussion}

In the following, we present simulation data obtained with a DM-EMC approach,
developed from our well-proven semiclassical EMC simulation tool by
implementing incoherent tunneling based on localized wavefunctions and a
self-consistent pure dephasing model, as described in Section \ref{sec:level2}%
. Pauli blocking is considered for all scattering mechanisms based on a
rejection technique \cite{1985Lugli_excl}, and non-equilibrium phonon
distributions (``hot phonons'') are explicitly taken into account
\cite{2006ApPhL..88f1119L}. For electron-electron interactions, an advanced
model including screening effects in random phase approximation as well as
electron spin is used \cite{2010JAP...107a3104J}. For impurity scattering,
screening is considered based on the modified single subband model
\cite{2010JAP...107a3104J,2006ApPhL..89u1115L}, and for LO phonon scattering,
screening is also taken into account \cite{ezhov2016influence}. Importantly,
our approach accounts for the coupling of the carrier transport to the optical
intensity evolution, allowing for a self-consistent simulation of
photon-induced electron transport and lasing
\cite{2010ApPhL..96a1103J,jirauschek2010monte,matyas2011photon}. Iterative
DM-EMC and Schr\"{o}dinger-Poisson simulations are performed until the
electron transport and lasing field converge to steady state.

\begin{figure}[ptb]
\includegraphics{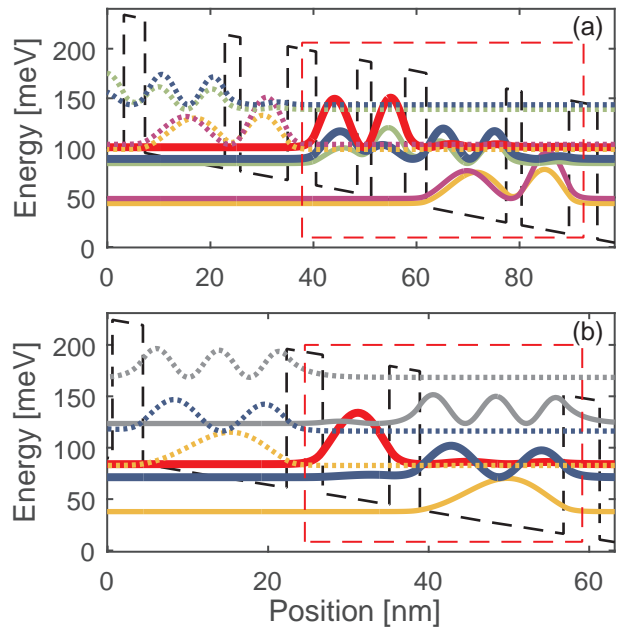}
\caption{Conduction band profiles and probability densities of the
investigated (a) four-well and (b) two-well THz QCL designs at a bias of
$10\,\mathrm{kV}/\mathrm{cm}$ and $13\,\mathrm{kV}/\mathrm{cm}$, respectively.
The main upper and lower laser level are marked by thick lines. The rectangles
denote a single QCL\ period.}%
\label{fig2}%
\end{figure}

We apply the DM-EMC approach to two different THz QCL designs, the four-well
LO phonon depopulation structure already shown in Fig.\thinspace\ref{fig1}
where we focus on the design G951 with a sheet doping density of
$1.2\times10^{10}\,\mathrm{cm}^{-2}$,\cite{benz:apl:2007} and a two-well
photon-phonon structure \cite{scalari2010broadband}. The conduction band
diagrams of the two structures at lasing bias, obtained with a tight-binding
Schr\"{o}dinger-Poisson approach, are shown in Fig.\thinspace\ref{fig2}.

We routinely consider four QCL modules, assuming periodic boundary conditions
for the first and last one \cite{2001ApPhL..78.2902I}, and simulate the
coupled evolution of the carrier transport and optical cavity field over
$70\,\mathrm{ps}$ to ensure convergence to steady state, using an ensemble of
$10,000$ electrons. As pointed out in Section \ref{sec:level1}, our main
motivation behind DM-EMC is to increase the reliability and versatility of
conventional EMC without sacrificing its relative computational efficiency.
Our Fortran implementation of DM-EMC requires about $25$ minutes on a single
thread of an Intel Xeon X5660 processor with a clock speed of
$2.8\,\mathrm{GHz}$ for both the two- and the four-well design. The required
memory is about $100\,\mathrm{MB}$ for the two-well and $150\,\mathrm{MB}$ for
the four-well design per bias point, allowing parallel simulations of many
bias points on a high performance server. This computational load is dominated
by the evaluation of the scattering processes, and thus comparable to that of
conventional EMC simulations.

\subsection{Current-voltage characteristics}

\begin{figure}[ptb]
\includegraphics{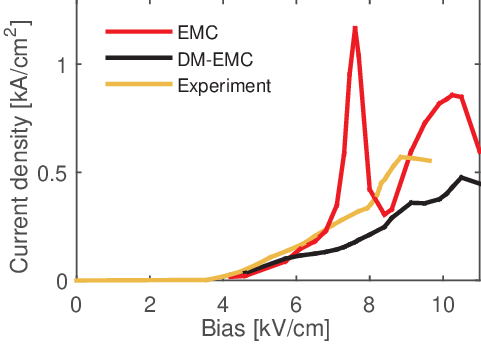}
\caption{Comparison of simulated and experimental current-voltage
characteristics for the four-well THz design at a lattice temperature of
$77\,\mathrm{K}$. Shown are simulation results obtained with EMC and DM-EMC,
and experimental data extracted from Ref.\thinspace\onlinecite{benz:apl:2007}%
.}%
\label{fig3}%
\end{figure}

In Fig.\thinspace\ref{fig3}, the DM-EMC result for the current-voltage
characteristics of the four-well THz QCL design is compared to semiclassical
EMC results and available experimental data. DM-EMC shows good overall
agreement with experiment. The computed peak current density is
$0.48\,\mathrm{kA}/\mathrm{cm}^{2}$ at a bias of $10.5\,\mathrm{kV}%
/\mathrm{cm}$, as compared to $0.57\,\mathrm{kA}/\mathrm{cm}^{2}$ at
$9\,\mathrm{kV}/\mathrm{cm}$ for the experiment. The slight deviations between
simulation and measurement can partly be attributed to growth deviations of
the experimental structure \cite{benz:apl:2007}. Furthermore, the exact amount
of cavity loss is somewhat uncertain, since theoretical values extracted from
waveguide modeling deviate significantly from experimental results obtained
for similar QCL structures \cite{benz:apl:2007}. For our simulations, we
assume waveguide and mirror power loss coefficients of $4\,\mathrm{cm}^{-1}$
and $3\,\mathrm{cm}^{-1}$, respectively, and a field confinement factor close
to $1$ \cite{benz:apl:2007,ban2006terahertz,2005JAP....97e3106K}. The peak
current density obtained by EMC in the lasing region is $0.86\,\mathrm{kA}%
/\mathrm{cm}^{2}$, significantly surpassing the experimental result.
Furthermore, EMC produces a distinct spurious current spike of
$1.17\,\mathrm{kA}/\mathrm{cm}^{2}$ at $7.6\,\mathrm{kV}/\mathrm{cm}$, which
is almost four times higher than the experimental value of $0.31\,\mathrm{kA}%
/\mathrm{cm}^{2}$ at this bias. Such current spikes are well known artifacts
of semiclassical EMC simulations
\cite{2003ApPhL..83..207C,2010PhyE...42.2628M,2007JAP...101h6109J}, emerging at biases in the
vicinity of small anticrossings. This scenario typically occurs for pairs of
coupled states extending across thick injection barriers, which would in the
semiclassical picture enable instantaneous electron tunneling across the
barrier without resistance \cite{2005JAP....98j4505C}. This situation is
illustrated in Fig.\thinspace\ref{fig1}(a), where the corresponding extended
wavefunctions are marked by bold lines. A suitable description of the
tunneling transport is provided by the density matrix formalism of Section
\ref{sec:tun}, combined with the corresponding tight-binding wavefunctions as
shown in Fig.\thinspace\ref{fig1}(b). Here, tunneling is mediated by the
coherent interaction between the left- and right-localized state, dampened by
dephasing processes which can be modeled as described in Section
\ref{sec:deph}.

\begin{figure}[ptb]
\includegraphics{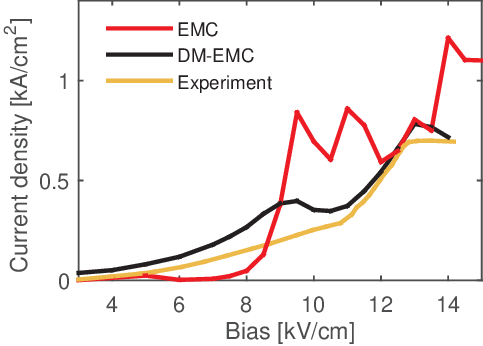}
\caption{Comparison of simulated and experimental current-voltage
characteristics for the two-well THz design at a lattice temperature of
$10\,\mathrm{K}$. Shown are simulation results obtained with EMC and DM-EMC,
and experimental data extracted from Ref.\thinspace\onlinecite
{scalari2010broadband}.}%
\label{fig4}%
\end{figure}

For further validation of DM-EMC, we have applied this approach to a two-well
photon-phonon THz QCL design. Figure \ref{fig4} contains the obtained
current-voltage characteristics, along with the corresponding EMC and
experimental results. For the simulation, we have assumed a total resonator
loss coefficient of $15\,\mathrm{cm}^{-1}$, and again a field confinement
factor close to $1$ \cite{scalari2010broadband,2005JAP....97e3106K}. The
DM-EMC result agrees well with the experimental data, with an additional bump
at around $9\,\mathrm{kV}/\mathrm{cm}$. Notably, such a feature has also been
observed in NEGF simulations of this structure at the same bias, where it is
attributed to tunneling through two barriers \cite{Wacker_JSTQE}. The EMC
simulation produces too low currents up to a bias of $8\,\mathrm{kV}%
/\mathrm{cm}$ and too high values above, caused by several extended current
spikes which are due to narrow anticrossings of coupled states extending
across the thick injection barrier.

\subsection{\label{sec:rates}Dephasing rates and subband electron distributions}

\begin{figure}[ptb]
\includegraphics{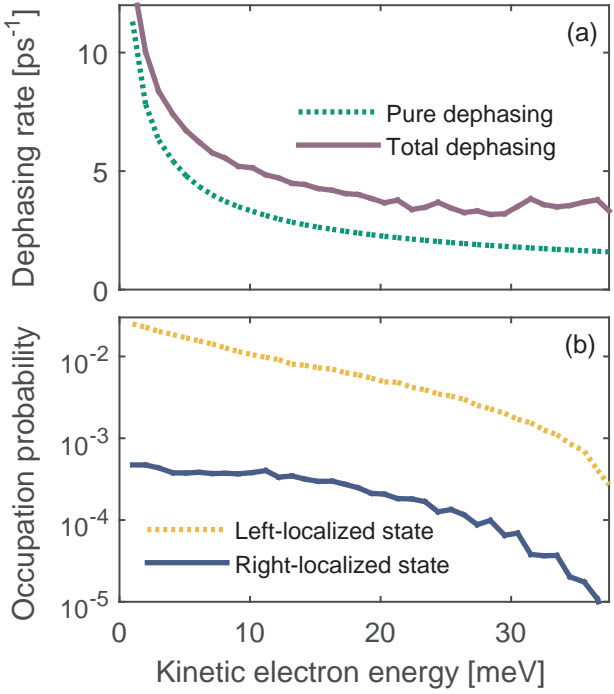}
\caption{(a) Pure and total dephasing rate as a function of kinetic electron
energy for the resonant tunneling transition across the thick barrier of the
four-well QCL design at a bias of $7.6\,\mathrm{kV}/\mathrm{cm}$ and lattice
temperature of $77\,\mathrm{K}$. (b) Occupation probability of the involved
states, marked by bold lines in Fig.\thinspace\ref{fig1}(b).}%
\label{fig5}%
\end{figure}

Figure \ref{fig5}(a) contains the pure and total dephasing rates as a function
of kinetic electron energy $\varepsilon$ for the coupled state doublet
mediating the tunneling transport across the injection barrier of the
four-well design, marked in Fig.\thinspace\ref{fig1}(b) by bold lines. The
pure dephasing is obtained as a sum of the interface roughness and the
impurity contribution, Eqs.\thinspace(\ref{gif}) and (\ref{gimp}). The total
dephasing rate additionally contains the lifetime broadening contribution,
computed based on the outscattering rate to states in other subbands
\cite{2009JAP...105l3102J}. The occupation probabilities $f\left(
\varepsilon\right)  $ of the two subbands involved in the tunneling transition
are shown in Fig.\thinspace\ref{fig5}(b)\ on a logarithmic scale.

In one-dimensional DM QCL simulation approaches
\cite{2009PhRvB..80x5316K,2010PhRvB..81t5311D,2010NJPh...12c3045T,dinh2012extended,tzenov2016time}%
, the kinetic electron energy is not explicitly taken into account, and
dephasing between two subbands $i$ and $j$ is often considered by an effective
rate $\gamma_{ij}=\left(  \gamma_{i}+\gamma_{j}\right)  /2+\gamma_{ij}%
^{\prime}$, where $\gamma_{i}$ and $\gamma_{j}$ are the inverse electron
lifetimes in subbands $i$ and $j$. The pure dephasing component $\gamma
_{ij}^{\prime}$ is typically treated as an empirical or fitting parameter
\cite{2009PhRvB..80x5316K,2010PhRvB..81t5311D,tzenov2016time,schrottke2010analysis,2005JAP....98j4505C}%
, but can also be obtained from Ando's model in Section \ref{sec:deph} by
averaging over the inversion between the corresponding subbands
\cite{2009JAP...106f3115N,freeman2016self}. Assuming in-plane isotropy, we
obtain with the in-plane wavevector $\mathbf{k}$, the kinetic electron energy
$\varepsilon=\hbar^{2}k^{2}/\left(  2m^{\ast}\right)  $ and $k=\left|
\mathbf{k}\right|  $%
\begin{align}
\gamma_{ij}^{\prime} &  =\frac{\sum_{\mathbf{k}}\gamma_{ij,\mathbf{k}}%
^{\prime}\left|  f_{i,\mathbf{k}}-f_{j,\mathbf{k}}\right|  }{\sum_{\mathbf{k}%
}\left|  f_{i,\mathbf{k}}-f_{j,\mathbf{k}}\right|  }\nonumber\\
&  \approx\frac{\int k\mathrm{d}k\gamma_{ij,k}^{\prime}\left|  f_{i,k}%
-f_{j,k}\right|  }{\int k\mathrm{d}k\left|  f_{i,k}-f_{j,k}\right|
}\nonumber\\
&  =\frac{\int\mathrm{d}\varepsilon\gamma_{ij}^{\prime}\left(  \varepsilon
\right)  \left|  f_{i}\left(  \varepsilon\right)  -f_{j}\left(  \varepsilon
\right)  \right|  }{\int\mathrm{d}\varepsilon\left|  f_{i}\left(
\varepsilon\right)  -f_{j}\left(  \varepsilon\right)  \right|  }.\label{gamav}%
\end{align}
For the case depicted in Fig.\thinspace\ref{fig5}, Eq.\thinspace(\ref{gamav})
yields a pure dephasing rate $\gamma_{ij}^{\prime}=4.62\,\mathrm{ps}^{-1}$ and
a lifetime broadening contribution of $1.74\,\mathrm{ps}^{-1}$, resulting in
$\gamma_{ij}=6.36\,\mathrm{ps}^{-1}$. Next, we investigate dephasing in the
four- and the two-well QCL design at the lasing biases considered in
Fig.\thinspace\ref{fig2}. For the pair of tunneling states with the narrowest
anticrossing, we obtain $\gamma_{ij}^{\prime}=2.04\,\mathrm{ps}^{-1}$,
$\gamma_{ij}=2.69\,\mathrm{ps}^{-1}$ for the four-well structure, and
$\gamma_{ij}^{\prime}=2.02\,\mathrm{ps}^{-1}$, $\gamma_{ij}=3.82\,\mathrm{ps}%
^{-1}$ for the two-well design. These obtained values of $\gamma_{ij}^{\prime
}$ are consistent with the expected pure dephasing times\ $\left(  \gamma
_{ij}^{\prime}\right)  ^{-1}\sim0.3..1\,\mathrm{ps}$, deduced from
measurements and fits to experimental data
\cite{freeman2016self,2009PhRvB..80x5316K,2010PhRvB..81t5311D,2005JAP....98j4505C}%
. For the main lasing transition, the dephasing rates are $\gamma_{ij}%
^{\prime}=0.24\,\mathrm{ps}^{-1}$, $\gamma_{ij}=2.37\,\mathrm{ps}^{-1}$ for
the four-well structure, and $\gamma_{ij}^{\prime}=0.75\,\mathrm{ps}^{-1}$,
$\gamma_{ij}=3.17\,\mathrm{ps}^{-1}$ for the two-well design, corresponding to
full width at half-maximum Lorentzian gain bandwidths $\gamma_{ij}/\pi$ of
$0.76\,\mathrm{THz}$\ and $1.01\,\mathrm{THz}$, respectively. As expected, for
the lasing transitions pure dephasing has a much smaller impact than for the
tunneling transport across thick barriers \cite{2010PhRvB..81t5311D},
justifying the use of lifetime broadening approaches to calculate spectral
gain bandwidths in QCLs \cite{2009JAP...105l3102J}. One reason is that the
pure dephasing contribution $\gamma_{ij}^{\prime}$\ tends to become small if
there is a considerable overlap between the corresponding wavefunctions, e.g.,
for vertical lasing transitions, since then the matrix elements in
Eq.\thinspace(\ref{gamp}) partially cancel \cite{2010PhRvB..81t5311D}. With
sheet doping densities of $1.2\times10^{10}\,\mathrm{cm}^{-2}$ and
$1.5\times10^{10}\,\mathrm{cm}^{-2}$ for the four- and two-well design,
respectively, and assuming typical interface roughness parameters of
$\Delta=0.12\,\mathrm{nm}$, $\Lambda=10\,\mathrm{nm}$
\cite{2008ApPhL..92h1102N}, the pure dephasing rates given above are dominated
by impurity scattering, which contributes at least 70\% to $\gamma
_{ij}^{\prime}$. Additionally, we have verified that an inclusion of LO
phonons would have modified above values of $\gamma_{ij}^{\prime}$ by less
than 4\%, justifying the omission of LO phonon contributions for the
evaluation of pure dephasing in such types of QCL designs
\cite{freeman2016self}. Instead of computing the lifetime broadening
contribution $\left(  \gamma_{i}+\gamma_{j}\right)  /2$ from the inverse
electron lifetimes $\gamma_{i}$ and $\gamma_{j}$ in subbands $i$ and $j$,
respectively, an alternative approach would be to take the corresponding term
$(\gamma_{i,\mathbf{k}}+\gamma_{j,\mathbf{k}})/2$ \ from Eq.\thinspace
(\ref{gijk}), and average it in the same way as $\gamma_{ij,\mathbf{k}%
}^{\prime}$, by using Eq.\thinspace(\ref{gamav}). This leads to different
results for transitions between subbands with non-identical electron
temperatures or even non-thermal kinetic energy distributions. However, for
the examples discussed above, both approaches yield similar values for the
total linewidth $\gamma_{ij}$ to within 10\%.

\begin{figure}[ptb]
\includegraphics{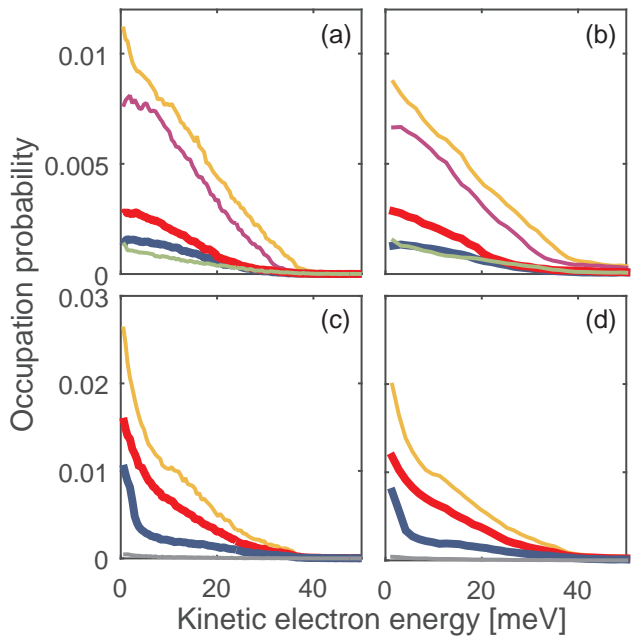}
\caption{Subband electron distributions at lasing bias for the four-well QCL
design at operating temperatures of (a) $10\,\mathrm{K}$ and (b)
$147\,\mathrm{K}$, and for the two-well design at (c) $10\,\mathrm{K}$ and (d)
$125\,\mathrm{K}$. The applied bias fields and line coding of the subbands are
as in Fig.\thinspace\ref{fig2}(a) and (b), respectively.}%
\label{fig6}%
\end{figure}

The electron distributions of the localized tunneling states shown in
Fig.\thinspace\ref{fig5}(b) are for a bias of $7.6\,\mathrm{kV}/\mathrm{cm}$,
where a narrow anticrossing of these states occurs (see Fig.\thinspace
\ref{fig1}). Here, almost all electrons in this subband doublet are localized
to the left of the barrier, which is the opposite scenario as assumed in the
semiclassical picture, where instantaneous tunneling results in equal
population distributions to the left and the right of the barrier
\cite{2005JAP....98j4505C}. This observation is consistent with the breakdown
of semiclassical EMC at this bias point, leading to the current spike shown in
Fig.\thinspace\ref{fig3}. In Fig.\thinspace\ref{fig6}, the subband electron
distributions of the four- and the two-well design are displayed at lasing
bias for low ($10\,\mathrm{K}$) and the maximum experimentally achieved
($147\,\mathrm{K}$ and $125\,\mathrm{K}$%
\ \cite{benz:apl:2007,scalari2010broadband}, respectively) operating
temperatures. For the four-well structure, there are considerably less
electrons in the right-localized tunneling state [the third highest occupied
level in Fig.\thinspace\ref{fig6}(a)\thinspace and (b)] than in the closely
aligned left-localized state (the highest occupied level), similarly as for
the case shown in Fig.\thinspace\ref{fig5}(b). Again, the current density
obtained with EMC at the corresponding bias point of $10\,\mathrm{kV}%
/\mathrm{cm}$ in Fig.\thinspace\ref{fig3} is too high since the piling up of
electrons behind the thick barrier is not contained in the semiclassical
description. By contrast, for the two-well design the occupation in both
tunneling states [the two highest occupied levels in Fig.\thinspace
\ref{fig6}(c)\thinspace and (d)] is comparable, and the EMC result for the
current density at the corresponding bias of $13\,\mathrm{kV}/\mathrm{cm}$ in
Fig.\thinspace\ref{fig4} agrees well with the DM-EMC simulation. Equivalent
electronic subband temperatures can be extracted from the expectation values
of the kinetic electron energy \cite{shi2014nonequilibrium,ezhov2016influence}%
. These are in the range of $121\,\mathrm{K}-145\,\mathrm{K}$ [Fig.\thinspace
\ref{fig6}(a)], $160\,\mathrm{K}-199\,\mathrm{K}$ [Fig.\thinspace
\ref{fig6}(b)], $116\,\mathrm{K}-127\,\mathrm{K}$ [Fig.\thinspace
\ref{fig6}(c)], and $154\,\mathrm{K}-165\,\mathrm{K}$ [Fig.\thinspace
\ref{fig6}(d)], significantly exceeding the lattice temperature (which
corresponds to the operating temperature at low duty cycles). This is the well
known hot electron effect, which has also been experimentally observed for
resonant phonon THz QCLs \cite{2005ApPhL..86k1115V}. We note that the subband
electron distributions deviate, in part, significantly from heated Maxwellian
distributions. In particular, the lower laser level distribution of the
two-well structure [third highest occupied level in Fig.\thinspace
\ref{fig6}(c)\thinspace and (d)] is highly non-Maxwellian, since its energetic
distance of $33.5\,\mathrm{meV}$ to the ground level is below the LO phonon
energy of $36\,\mathrm{meV}$. Consequently, the LO\ phonon depopulation
channel is blocked for electrons at the bottom of the lower laser level.
While, as mentioned above, the piling up of electrons behind thick tunneling
barriers can only be modeled by considering quantum coherence effects, hot
electrons and non-Maxwellian subband carrier distributions are also adequately
described by semiclassical simulations
\cite{2001ApPhL..78.2902I,2004ApPhL..84..645C,2008pssc..5..221J,iotti2010impact,shi2014nonequilibrium}%
.

\section{\label{sec:level4}Conclusion}

In conclusion, we have presented a DM-based method to self-consistently
include incoherent tunneling into the EMC framework for QCL\ simulations. The
resulting DM-EMC approach maintains the strengths of EMC, such as its relative
computational efficiency and the inclusion of intercarrier scattering as well
as carrier-light coupling, while partially curing the main shortcoming of
semiclassical QCL modeling techniques, i.e., the omission of quantum coherence
effects. Out of these, tunnel coupling between state doublets spanning thick
barriers plays an eminent role. For narrow anticrossings of the tunneling
states, as especially occur in THz QCL designs with thick injection barriers,
a semiclassical treatment can even break down completely, which manifests
itself in the emergence of an artificial spike in the current-voltage
characteristics. By a self-consistent implementation of tunnel dephasing based
on Ando's model, we have in our approach eliminated the need for empirical
dephasing times. We have validated our simulation scheme against experimental
data for a two- and a four-well THz QCL\ design, clearly demonstrating the
superiority of DM-EMC over conventional EMC for those structures. The
developed approach is not only relevant for steady-state QCL\ simulations, but
will also be useful for Maxwell-Bloch-type modeling of the QCL dynamics where
a coupling to steady-state carrier transport simulations can yield the
required scattering and dephasing rates
\cite{tzenov2016time,tzenov2017analysis}.%

\begin{acknowledgments}
This work was supported by the German Research Foundation (DFG) within the
Heisenberg program (JI 115/4-1) and under DFG Grant No. JI 115/9-1.
\end{acknowledgments}

\end{document}